\documentclass[preprint,floatfix,aps,showpacs,showkeys]{revtex4}
\usepackage{epsfig,latexsym}
\usepackage{color}
\usepackage{xcolor}
\usepackage{amssymb,amsmath}

\newcommand {\bea}{\begin{eqnarray}}
\newcommand {\eea}{\end{eqnarray}}
\newcommand {\be}{\begin{equation}}
\newcommand {\ee}{\end{equation}}

\begin{document}
\def\Journal#1#2#3#4{{#1} {\bf #2}, #3 (#4)}
\def\RPP{{Rep. Prog. Phys.}}
\def\PRC{{Phys. Rev. C}}
\def\PRD{{Phys. Rev. D}}
\def\ZPA{{Z. Phys. A}}
\def\NPA{{Nucl. Phys. A}} 
\def\JPG{{J. Phys. G }}
\def\PRL{{Phys. Rev. Lett.}}
\def\PR{{Phys. Rep.}}
\def\PLB{{Phys. Lett. B}}
\def\AP{{Ann. Phys (N.Y.)}}
\def\EPJA{{Eur. Phys. J. A}}
\def\NP{{Nucl. Phys.}}  
\def\RMP{{Rev. Mod. Phys.}}
\def\IJMPE{{Int. J. Mod. Phys. E}}
\def\AJ{{Astrophys. J.}}
\def\AJL{{Astrophys. J. Lett.}}
\def\AA{{Astron. Astrophys}}
\def\ARAA{{Annu. Rev. Astron. Astrophys}}
\def\MPLA{{Mod. Phys. Lett. A}}
\def\ARNPS{{Annu. Rev. Nuc. Part. Sci}}
\def\LRR{{Living. Rev. Relativity}}
\def \PTP{{Prog. Theor. Phys.}}
\def\ZPC{{Z. Phys. C}}

\title{IMPRINT OF NUCLEON RADIUS ON THE SLOWLY ROTATING NEUTRON STAR PROPERTIES}

\author{Suparti}
\author{A. Sulaksono}\email{anto.sulaksono@sci.ui.ac.id}
\author{T. Mart}

\affiliation{Departemen Fisika, FMIPA, Universitas Indonesia, Depok 16424, Indonesia }

\begin{abstract}
We have studied the effect of free space nucleon radius on the 
nuclear matter and slowly rotating neutron star properties
within a relativistic mean field model with the inclusion of omega-rho mixing 
nonlinear term. We use the same density dependent radius as in our 
previous work~\cite{MS2013}. However, in the present work we 
utilize the parameter sets with more proper symmetric 
nuclear matter (SNM) predictions for a number of different 
nucleon radii. To this end, the isoscalar parameters in each 
parameter set and the cutoff parameter $\beta$ are adjusted 
by fitting the corresponding SNM 
properties at sub-saturation density to the prediction of 
IUFSU parameter set~\cite{Fattoyev2010}. 
We have obtained that the assumption of 
composite nucleons with free space radius 
$r \le 0.83$ fm is still compatible with the currently acceptable 
nuclear matter (NM) properties beyond the saturation density. 
We have also found that the effect of free space nucleon radius could 
be imprinted in the radius, moment of inertia, and crust properties 
of the slowly rotating neutron star.
\end{abstract} 

\keywords{Proton radius, neutron star, moment of inertia}
\pacs{26.60.c, 21.65.f, 13.40.Gp, 14.20.Dh}

\maketitle
 \section{INTRODUCTION}
\label{sec_intro}

Recent analysis on the mass distribution of a number of pulsars with 
secure mass measurement has confirmed that $M \sim 2.1 M_\odot$ provides 
an established lower bound value on the maximum mass $(M_{\rm max})
$ of neutron star (NS) \cite{KKYT2013}. The evidence of massive NS 
with accurate measurement was obtained through the recent observations 
of J1614-2230 pulsar from the Shapiro delay \cite{Demorest10} and  
J0348+0432 pulsar from the gravitational redshift of its white dwarf 
companion \cite{Antoniadis13}.  The corresponding J1614-2230 and 
J0348+0432 pulsar masses are  $1.97~\pm~0.04~M_\odot$ and 
$2.01~\pm~0.04~M_\odot$, respectively. However, up to now, the 
analysis methods used to extract the NS radii from observational 
data still suffer from  high uncertainty \cite{MCM2013,Bog2013,
Gui2013,LS2013,Leahy2011,Steiner2010,Stein2013,Sule2011,Ozel2016}. 
Within the framework of general relativity and assuming that 
baryons are point-like particles (see Ref.~\cite{Miller2016} and 
references therein) theoretical prediction of a $1.5 M_\odot$ NS 
yields a radius between 10-15 km. At this stage we need to note
that although the large NS radii are not totally excluded 
by astrophysical observations, they are incompatible with experimental 
data and standard many body calculations (see Ref.~\cite{Delsate2016} 
and the references therein). 

Uncertainties in the  equation of state (EOS) due to the poorly known density dependence of symmetry energy  $S(\rho)$ significantly affect the transition properties between the core and the crust of NS. Note that the symmetry energy $S(\rho)$ is the energy difference between the SNM and the pure neutron matter (PNM) at a fixed baryon density $\rho$. Although $S(\rho)$ is experimentally well determined at the saturation density $\rho_0$, the value of its slope $L$ at  $\rho_0$, which controls the properties of neutron-rich matter at the crust-core transition region, is still uncertain. The values of core-crust transition density $\rho_t$ and proton fraction $Y_p$ around $\rho_t$  are both linearly anti-correlated to $L$. However, the pressure  at the transition density, $P_t$, does not show a similar behavior (see Ref.~\cite{PFH2014} and references therein). Note also that a recent systematic study on the corelation of NS radii with $L$ and the slope of NM incompressibility in a wide range of NS EOS has been just performed and reported in Ref.~\cite{CR2016}. Further discussions of the correlations of NS core-crust properties with $L$  predicted by different methods can be also found in Ref.~\cite{PSAP2016} and the references therein. Furthermore, the uncertainties in NS radii and crust properties due to the limited knowledge of EOS have been also discussed recently in Ref.~\cite{Fortin2016}. 

Concerning the NS crust, a recent study by Delsate {\it et al.}
\cite{Delsate2016} has shown the necessity to treat all regions 
of NS within the same nuclear model to ensure that the EOS is 
thermodynamically consistent. Furthermore, they have suggested 
that the role of neutron superfluidity in core could be more 
important than before, because they have found that the neutron 
superfluidity in the crust of an NS does not carry 
enough angular momentum to explain the giant frequency glitches 
in the Vela pulsar. Piekarewicz $\it et. al.$ \cite{PFH2014} also explored the possibility that uncertainties in the NS EOS can provide a sufficient flexibility for the
construction of models that predict a large  thickness and moment of inertia of the NS crust. They obtained  fractional moments of inertia as large as 7 $\%$ for the NSs with masses below 1.6  $M_\odot$. They also found that if the neutron-skin thickness of $^{208}$Pb were in the range of 0.20 - 0.26 fm,  sufficiently large transition pressures  could be generated to explain the large Vela glitches without requiring additional angular-momentum beyond that confined to the solid crust \cite{PFH2014}.

On the other hand, it has been reported in 
Ref.~\cite{Raithel2016} that if the equation of state is 
trusted up to the nuclear saturation density, measurement 
of the moment of inertia of PSR J0737-3039A will 
place absolute bounds on the corresponding radius within $\pm 1$ km. 
However, it seems that the authors of Ref.~\cite{Raithel2016} did not analyze the full range of possible EOSs. Steiner {\it et al.} \cite{SGFN2015} performed a systematic assessment of models for the EOS of dense matter in the context of recent neutron star mass and radius measurements. They have shown that the currently available neutron star mass and radius measurements might provide a strong constraint on the moment of inertia, tidal deformabilities, and crust thicknesses. Furthermore, a measurement of the moment of inertia of PSR J0737-3039A with a 10 $\%$ error, without any other information from observations, will constrain the EOS over a range of densities to within 50 - 60 $\%$. Therefore, the radius determination solely from the corresponding measurements can be less stringent. The result of   the moment of inertia of PSR J0737-3039A measurement can be expected within the next five years.

The proton charge radius extracted from recent muonic hydrogen Lamb shift 
measurements is significantly smaller than that extracted from atomic 
hydrogen and electron scattering measurements~\cite{MS2013}. 
The discrepancy has become 
known as the proton radius puzzle (see, e.g., Refs.~\cite{Higin2016,Madox2016} 
and references therein for the current status of this puzzle). 
So far, in the standard picture of NM and NSM the baryons have been assumed to be point particles. 
A number of studies of the nucleon radius effect on NM and 
NS properties have been previously performed, e.g., 
in Refs.~\cite{MS2013,Kouno1996,Rischke1991}. However, a more 
quantitative study of the  effect of nucleon radius on NM 
and NS properties by comparing the predictions 
with experimental and observational data has not yet been done. 
Furthermore, by considering the rapid progress in observational 
and theoretical studies in the field of NM and 
NS it is expected that  the uncetainty in NS EOS can be significantly reduced 
within the next few years. Therefore, it is obvious that a quantitative 
study of the impact of the nucleon radius on NM 
and NS properties becomes timely and necessary. 

In this paper, we report on the result of our study on 
the effect of free space nucleon radius on the NM 
and slowly rotating NS properties within 
a relativistic mean field (RMF) model with the inclusion of omega-rho 
mixing nonlinear term. In the calculation we use the same density dependent 
radius as in our previous work~\cite{MS2013}. However, different 
from the previous work, where the IUFSU parameter set~\cite{Fattoyev2010} 
along with a fixed cutoff parameter $\beta$ were used, in the present 
work we generate a number of parameter sets for the RMF model with 
different values of free space nucleon radius $r$ by adjusting the 
parameters of isoscalar and cutoff $\beta$, so that the SNM properties at saturation density coincide 
with the ones predicted by the IUFSU parameter set~\cite{Fattoyev2010}.

In Sec.~\ref{formalism} we briefly discuss the formalism used to 
calculate the EOS  and the slowly rotating NS properties. In 
Sec.~\ref{RAD} we present the results and discuss their consequence. 
We will conclude our findings in Sec.~\ref{sec_conclu}.

\section{FORMALISM}
\label{formalism}
In this section we briefly discuss the formalism used to calculate 
the EOS of NM and NS matter with taking into account the 
nucleon excluded volume effect. We also present the formalism used 
to calculate the properties of slowly rotating NS. 

\subsection{Matters with nucleon excluded volume effect }

Based on the RMF model the energy density of matter consisting of 
leptons and composite nucleons with a radius $r$ can be written 
as~\cite{Kouno1996,Rischke1991,MS2013}
\begin{eqnarray}
\epsilon&=&\mathcal{A} [\epsilon_p^k + \epsilon_n^k] +\epsilon_e^k + \epsilon_{\mu}^k +\epsilon_M(\omega,\sigma,\rho)\nonumber\\
 &+& g_{\omega} \omega_0 (\rho_p+\rho_n)+{\textstyle \frac{1}{2}} g_{\rho} b_0 (\rho_p-\rho_n).
\label{eq:eden} 
\end{eqnarray}
In Eq.~(\ref{eq:eden})
 $\epsilon_M$ is the total energy density of $\sigma$, $\omega$ and $\rho$ mesons, including the $\sigma$, $\omega$ and $\rho$-$\sigma$ mixing 
nonlinear contributions, where the explicit form reads
\begin{eqnarray}
\epsilon_M &=&\epsilon_M^{\rm linear}+\frac{1}{3} b_2\sigma^3+ \frac{1}{4} b_3\sigma^4\nonumber\\&-&\frac{1}{4} c_3  \omega^4 - \Lambda_v g_\rho^2 g_\omega^2\rho^2 \omega^2 .
\label{eq:Mesedens}
\end{eqnarray}
The coupling constants of $\omega$-, $\sigma$-, and $\rho$-meson 
are denoted by $g_{\omega}$,  $g_{\sigma}$ and  $g_{\rho}$, respectively,
whereas  the corresponding meson fields are denoted by 
$\omega$,  $\sigma$ and  $\rho$, respectively. 
The excluded volume effect enters into the model 
through the normalization constant $\mathcal{A}$, which is 
defined as~\cite{Kouno1996,Rischke1991,MS2013}
\begin{equation}
\mathcal{A}\equiv\frac{1}{1+V_p \bar{\rho}_p+V_n \bar{\rho}_n} .
\end{equation}
Here $V_p$ and  $V_n$ denote the volumes of proton and neutron, 
respectively. The point particle kinetic energy of degenerate Fermi gas is
\begin{equation}
\epsilon_i^k= \frac{2}{{(2 \pi)}^3} \int d^3\vec{k} {(k^2 +m^{*~2}_i)}^{1/2}\theta(k-k_F). 
\end{equation}
In the case of non-interacting lepton the effective mass 
$m^{*}_i =  m_i$, whereas in the case of interacting nucleon  
$m^{*}_i=m_i-g_{\sigma} \sigma$, where for lepton $i= e,\mu$ and for nucleon $i=p,n$. 
Here, the proton (or neutron) and the corresponding scalar 
densities read  
\begin{eqnarray}
\rho_i= \mathcal{A} \bar{\rho}_i ~,~~
\rho_{s,i}= \mathcal{A} \bar{\rho}_{s,i}
\end{eqnarray}
where $\bar{\rho}_i$ and $\bar{\rho}_{s,i}$ are the the corresponding point 
particle densities.  To simplify the calculation, we assume that  $V_p \approx V_n \equiv V_N$ 
where $V_N = \frac{4}{3} \pi r^3$, with $r$ denotes the nucleon radius. The pressure of matter can be obtained from 
Eq.~(\ref{eq:eden}) by using the thermodynamics relation
\begin{equation}
P=\rho^2 \frac{d(\epsilon/\rho)}{d\rho}.
 \end{equation}
The proton or neutron chemical potential can be obtained from 
\begin{equation}
\mu_i=E^*_{F,i}+V_i P_i'+ g_{\omega} \omega_0 +\alpha_i \frac{1}{2} g_{\rho} b_0,
\label{eq:chemical}
\end{equation}
where $\alpha_i= 1$ for proton and $\alpha_i=-1$ for neutron.
Furthermore, $E^*_{F,i}$=
${(k_{F,i}^2 +m^{*2}_i)}^{1/2}$. In Eq.~(\ref{eq:chemical}) 
the function  $P_i'$ is defined as
\begin{eqnarray}
 P_i'=\frac{1}{12 \pi^2} \left[ E^*_{F,i} k_{F,i}(E^{*2}_{F,i} -\frac{5}{2} m^{*2}_i)
     +\frac{3}{2} m^{*4}_i ~\log\left(\frac{k_{F,i}+E^*_{F,i}}{+m^*_i}\right)\right].
\end{eqnarray}

Different from the quark meson coupling model, where the density dependent 
of nucleon radius  $r(\rho)$ can be directly obtained from the 
model~\cite{Panda2002,Aguirre2003}, in the RMF model one should use a 
phenomenological form of $r (\rho)$ proposed in Ref. \cite{MS2013}, i.e.,
\begin{equation}
r(\rho) = {r}\left[1+\beta {\left(\frac{\rho}{\rho_0}\right)}^2\right]^{-\gamma},
\label{rrho}
\end{equation}
where $\gamma$ is taken equal to 2 in our calculation, while $\rho=\rho_p+\rho_n$ and $\rho_0$ is the SNM saturation density. 
The physical reason of choosing the form of $r(\rho)$ and its relation 
to the quark meson coupling model has been already discussed in 
Ref.~\cite{MS2013}.  However, unlike in Ref~\cite{MS2013}, where 
a fixed cutoff $\beta$ and the IUFSU 
parameter set~\cite{Fattoyev2010} were used, in this work we generate 
different parameter sets of the RMF model with different values of $r$ by 
adjusting the isoscalar parameters and $\beta$, such that the SNM 
properties at sub-saturation density coincide with those predicted by 
the IUFSU parameter set~\cite{Fattoyev2010}. For the isovector 
parameter set we simply take the IUFSU parameter set. In this way 
we can get more reasonable SNM and PNM 
properties at saturation density $\rho_0$ than in the case of
our previous work \cite{MS2013}.

\begin{figure}
\epsfig{figure= 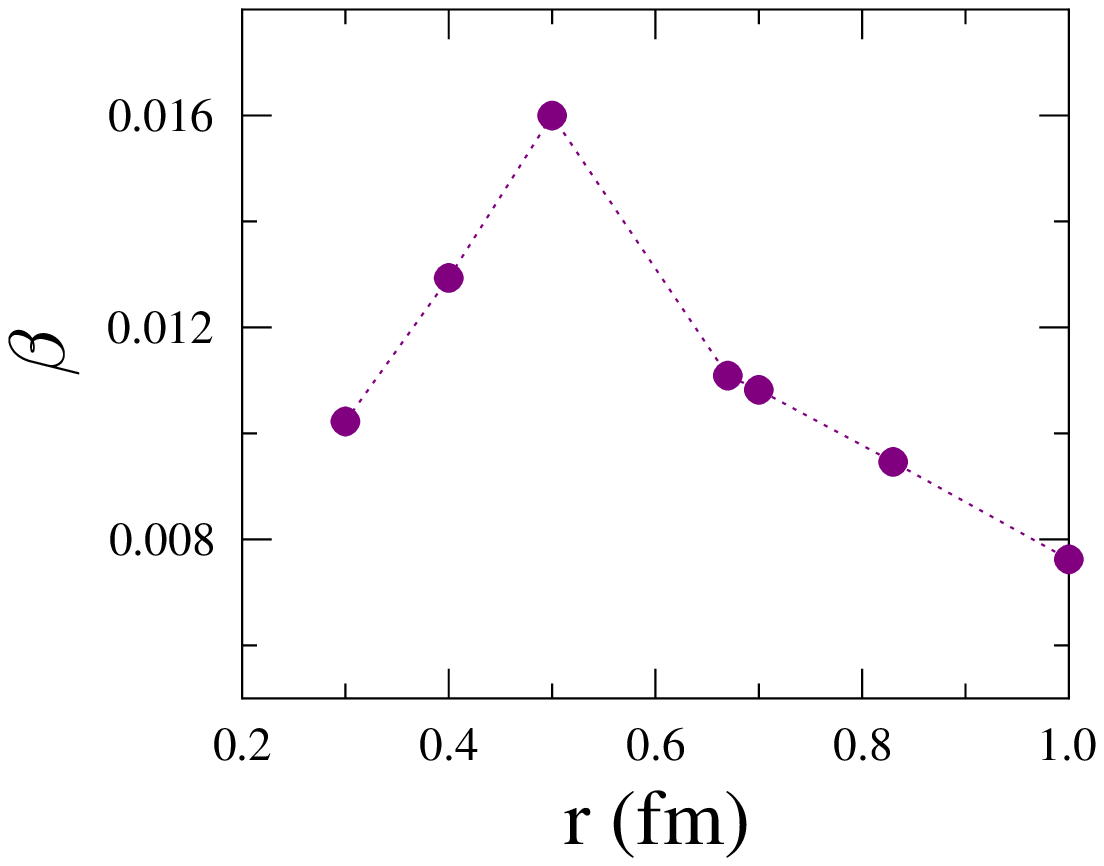, width=10cm}
\caption{(Color online) Cutoff parameter $\beta$ as a function of 
  the free space nucleon radius.}
\label{fig:betarn}
\end{figure}

The values of RMF parameters used in the present work are listed in 
Table ~\ref{tab:parset}.  Except for the value of $b_2$ in 
the case of $r=1$ fm, 
it is obvious that the SNM properties at saturation force 
the isoscalar parameters of the model, such as $g_\sigma$, 
$g_\omega$, $b_2$, $b_3$ and $c_3$, to increase, as the nucleon 
radius $r$ increases. It can be also seen in Fig.~\ref{fig:betarn} 
that there exists a rather unintuitive  relation between the nucleon 
radius $r$ and the cutoff parameter $\beta$,  where $\beta$ increases 
with $r$ but only up to $r= 0.5$ fm and decreases for $r \ge 0.5$ fm. 
This result indicates that for nucleons with $r \ge 0.5$ fm the 
density dependent $r (\rho)$ is less compressed as $r$ increases.  

\begin{table} 
\caption {RMF parameters for different values of the nucleon radius 
$r$ used in this work obtained from fit. The values of isovector 
parameter  were fixed to the values of the IUFSU parameter set 
($r=0$ fm), whereas the isoscalar parameters and $\beta$ were varied 
during the fit process. In the last line f (c) indicates that the 
parameters were fitted (fixed).}
\label{tab:parset}
\begin{ruledtabular}
\begin{tabular}{c c c c c c c c c }
$r$ (fm) & $g_\sigma$  & $g_\omega$  & $b_2$ (MeV) & $b_3$       & $c_3$       & $\beta$     & $g_\rho$    & $\Lambda_v$ \\
\hline
0        & 9.97     & 13.03    & $-1675.88$ & 0.48     &144.22    &-         &13.59     &0.046     \\
0.4      &9.95   &12.98  &$-1723.3$ & $-2.0$  &146.70 &0.013  &13.59  &0.046  \\
0.67     & 9.91     & 12.81    &$-1950.0$   &-15.324 &130.0     &0.011     &13.59     & 0.046    \\
0.83     &10.35   & 13.59 &$-1850.0$ & $-20.80$ &190.0   &0.009   &13.59   &0.046   \\
1.00     & 11.28 &  15.27&$-1720$.4&$-24.0$  &399.68 &0.0076 &13.59  &0.046  \\
c &f&f&f&f&f&f&c&c\\
\end{tabular}\\
\end{ruledtabular}
\end{table}

\subsection{Slowly rotating neutron stars }

To calculate the NS properties, we start with the line element  
of slowly rotating NS \cite{IOJ2015},
\bea
ds^2&=&-e^{2 \nu} dt^2 +e^{2 \lambda} dr+ r^2 (d \theta^2 + sin^2 \theta ~d\phi^2) \nonumber\\ &-& 2 \omega(r)~r^2~sin^2 ~\theta ~dt ~d\phi ,
\label{LEE}
\eea
where $\omega(r)$ accounts for the frame-dragging effect. From the 
definition of $\bar{\omega}(r) \equiv \Omega-\omega(r)$, it is apparent 
that the line element given in Eq.~(\ref{LEE}) is  only correct up 
to order of $\Omega$. This slowly rotating approximation means that the 
star retains its spherical geometry, since the centrifugal deformation 
is considered to be at the order of $\Omega^2$~\cite{IOJ2015}. Solving 
the Einstein equation by means of this metric and assuming that the NS matter 
is a perfect fluid, one can obtain the Tolman-Oppenhaimer-Volkoff (TOV) 
equation
\bea
\frac{dm}{dr} &=& 4 \pi \epsilon r^2,\nonumber\\
\frac{dp}{dr} &=& -G \frac{\epsilon m}{r^2}\left(1+\frac{p}{\epsilon}\right)\left(1+\frac{4 \pi r^3 p}{m}\right){\left(1-\frac{2Gm}{r}\right)}^{-1},
\label{TOV}
\eea
with the $\nu$ function in metric obeys the first order ordinary differential equation as
\be
\frac{d\nu}{dr} = G \frac{m +4 \pi r^3 p}{r(r-2 G m)},
\label{matric_NU}
\ee
and $\bar{\omega}$ function obeys the second order ordinary differential equation as
\be
\frac{1}{r^4}\frac{d}{dr}\left[r^4  e^{-\nu} {(1-\frac{2 G m}{r})}^{1/2}  \bar{\omega}\right]+\frac{4}{r^4}\left[\frac{d}{dr} e^{-\nu} {(1-\frac{2 G m}{r})}^{1/2}\right] \bar{\omega}=0 .
\label{omega}
\ee
Using the NS EOS $P=P(\epsilon)$ calculated by means of the 
method given in the previous sub-section, Eqs.~(\ref{TOV})-(\ref{omega}) 
can be numerically integrated by utilizing a standard method. 
Since at $r=R$ (the star radius), ${d\bar{\omega}}/{dr} = 
{ 6 G I \Omega}/{R^4}$, the moment of  inertia $I$ can be obtained 
from Eq.~(\ref{omega}).   

Note that to obtain the EOS of NS matter, 
we assume that the corresponding matter is neutral and obey the  $\beta$ 
stability condition. Furthermore, it is well known that the NS 
can be divided into two regions, i.e., the crust and core,
with different compositions, particle distributions, and density ranges. 
In the present work, we use the crust EOS calculated by Miyatsu 
$et ~al$.~\cite{MYN2013}, while for the core EOS we use the one 
obtained from the RMF model with the inclusion of omega-rho mixing 
nonlinear term (FSU type), which is generally formulated in the previous 
subsection. It is important to emphasize here that we still do not 
use a unified core and crust EOS. The effect of a unified core-crust 
EOS can be considered as a further correction in the crust properties. 
To determine the core-crust transition density, we use one of the standard methods i.e., random phase approximation (RPA) method  by assuming all involving particles in RMF model are point particles (see for example Ref.~\cite{CHP2003} for the details of the method) for all parameter sets used. Including the effect of free space radius of nucleon in the used RPA method might lead to a further correction in the crust properties. In this work we also exclude the exotic constituent of matter, such as hyperons, since their coupling constants are quite
uncertain. The global effect of this contribution can be expected to 
decrease the NS maximum mass and to affect the crust properties.

\section{RESULTS AND DISCUSSIONS}
\label{RAD}
In this section, we show that the SNM and PNM EOSs of the system 
consisting of composite nucleons with  $r \le 0.83$ fm, 
calculated by using the RMF model, are still compatible with the
currently acceptable NM properties. We also discuss 
the imprint effect of finite nucleon radius on the NM 
in the slowly rotating NS properties.

\subsection{Impact of the nucleon radius on the nuclear matter properties}

\begin{figure}
\epsfig{figure= 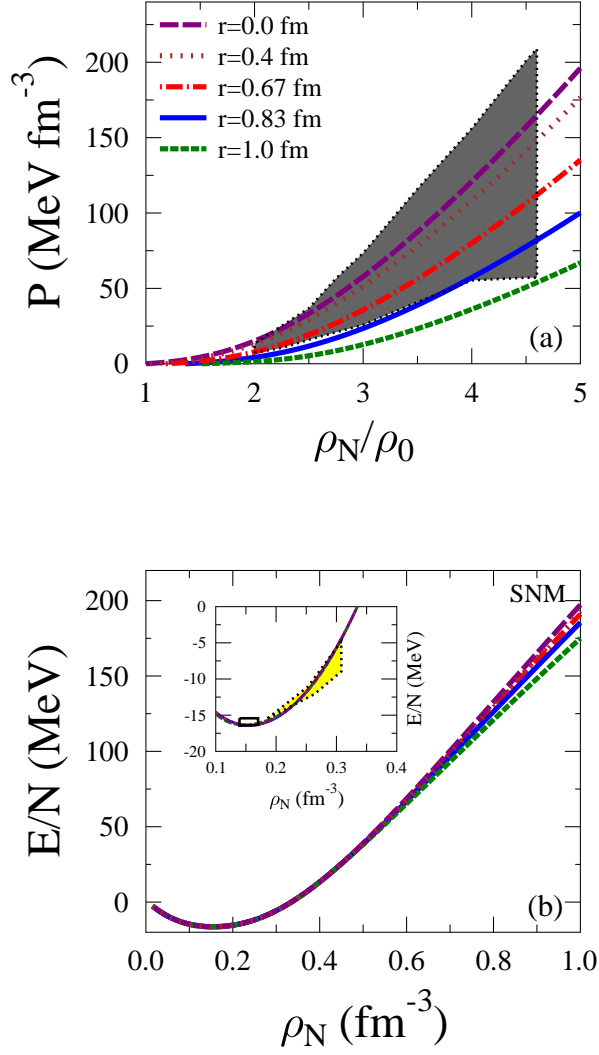, width=9cm}
\caption{(Color online) (a) Pressure as a function of the 
  ratio between nucleon and nuclear saturation densities. (b) Energy per particle as a function of the density around the normal density for the symmetric nuclear matter with different nucleon radii. Shaded area 
  in (a) corresponds to the heavy-ion experimental data taken from 
  Ref.~\cite{Daniel}. The shaded area in the inset of (b)  
  displays the constraint of nuclear matter EOS near twice the saturation density from the FOPI data~\cite{LFevre2016}.  }
\label{fig:snm}
\end{figure}

The binding energy per nucleon ($E/N$) as a function of the nuclear 
matter density along with the pressure as a function of the ratio 
between nucleon and nuclear saturation densities, i.e., $\rho_N$ and 
$\rho_0$, at moderate density region for SNM with $0< r< 1$ fm 
are shown in the upper and lower panels of Fig.~\ref{fig:snm}. 
It can be seen in Fig.~\ref{fig:snm}b that 
at $\rho_0$ the SNM $E/N$ obtained from all parameter sets, with 
different free space nucleon radii $r$, are in agreement with 
the experimental data and the prediction of IUFSU parameter set 
(in the figure, it is indicated by the curve with $r=0.0$ fm).
Furthermore,  up to twice the saturation density, they are also 
compatible with the result extracted from FOPI data~\cite{LFevre2016}. 
This result can be understood because 
we have fitted the isoscalar parameters of the corresponding parameter 
set with  $r\ne 0$ in order that the binding energies in the vicinity of 
$\rho_0$ are compatible with those predicted by the IUFSU parameter set. 
The difference in the binding energies starts to appear as 
$\rho_N$ $\gtrsim$ 0.7 fm$^{-3}$. However, it can be observed in 
Fig.~\ref{fig:snm}a that the effect of nucleon 
radius appears significantly to soften the EOS (the pressure-density 
relation) as $\rho_N$ $\gtrsim$ 2  $\rho_0$. To be more precise, 
the EOS becomes softer as the radius $r$ increases. The extracted 
SNM EOS from heavy ion experimental data of Danielewicz 
{\it et al.}~\cite{Daniel} provides a quite stringent upper-bound 
of $r$, i.e., $r \le$ 0.83 fm. 

\begin{figure}
\epsfig{figure= 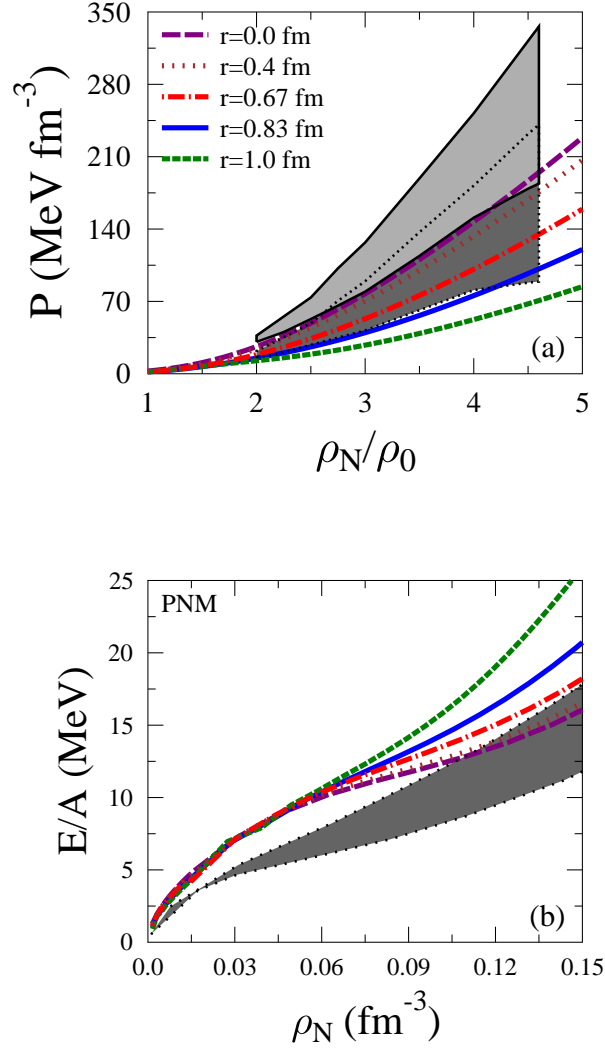, width=9cm}
\caption{(Color online) (a) Pressure as a function of the ratio 
  between nucleon and nuclear saturation densities. (b) Energy per particle as a function of the density at low density region 
  for pure neutron matter with different nucleon radii. Shaded area 
  in (a) corresponds to the heavy-ion experimental data based 
  on soft and stiff symmetry energies taken from Ref.~\cite{Daniel},
  whereas shaded area in (b) exhibits the  pure neutron matter 
  result of Ref.~\cite{KTHS2013}.}
\label{fig:pnm}
\end{figure}

The binding energy per neutron ($E/N$) 
as a function of the density at very low density along with the pressure 
as a function of the ratio between nucleon and nuclear saturation 
densities for pure neutron matter (PNM) with $0< r< 1$ fm are shown 
in the upper and lower panels of Fig.~\ref{fig:pnm}. 
Figure~\ref{fig:pnm}a clearly displays that only the EOS 
with $r \le 0.83$ fm would be compatible with the extracted soft 
PNM EOS of Danielewicz {\it et al.}~\cite{Daniel}. This result is 
clearly consistent with the one obtained in the case of SNM. 
However, as shown in Fig.~\ref{fig:pnm}b, 
the PNM binding energy at very low density obtained in the present 
work and the one obtained with $r=0.0$ fm (the IUFSU parameter set) 
seem to be somewhat stiffer than the recent result obtained from 
the chiral effective field theory calculation~\cite{KTHS2013}. 
We need to note here that the problem at very low density in 
the PNM originates from the isovector parameter used in the
calculation, instead from the excluded volume effect. 
In the present work, we have only used the parameter given 
in the IUFSU parameter set. We also note that the IUFSU PNM 
EOS at very low densities is not compatible with the one given
in Ref.~\cite{KTHS2013}. The problem in PNM EOS might be remedied 
by fitting the isovector parameters to the data given in 
Ref.~\cite{KTHS2013}. This will be the topic of our future 
investigation.  At this stage, we might conclude that 
the EOS of the system consisting of composite nucleons is still 
compatible with the currently acceptable NM properties. 

On the other hand, the existence of the upper-bound of nucleon radius 
would become very interesting if we relate it to the recent 
analyses on the proton charge radius extraction 
from the proton scattering data~\cite{Higin2016,Madox2016}. 
We note that these analyses are compatible with 
the recent muonic hydrogen result, i.e., 
$r_p = 0.84$ fm~\cite{Pohl2010,Antognini2013}. 
Therefore, the imprint of nucleon radius in the NM will
shed a new light on the proton radius puzzle.

\subsection{Impact of the nucleon excluded volume on the neutron star properties}

Here we discuss how the effect of free space nucleon radius is imprinted 
 in the slowly rotating NS properties, such as the mass-radius relation, 
the NS moment of inertia and the crust properties. 

\begin{figure}
\epsfig{figure= 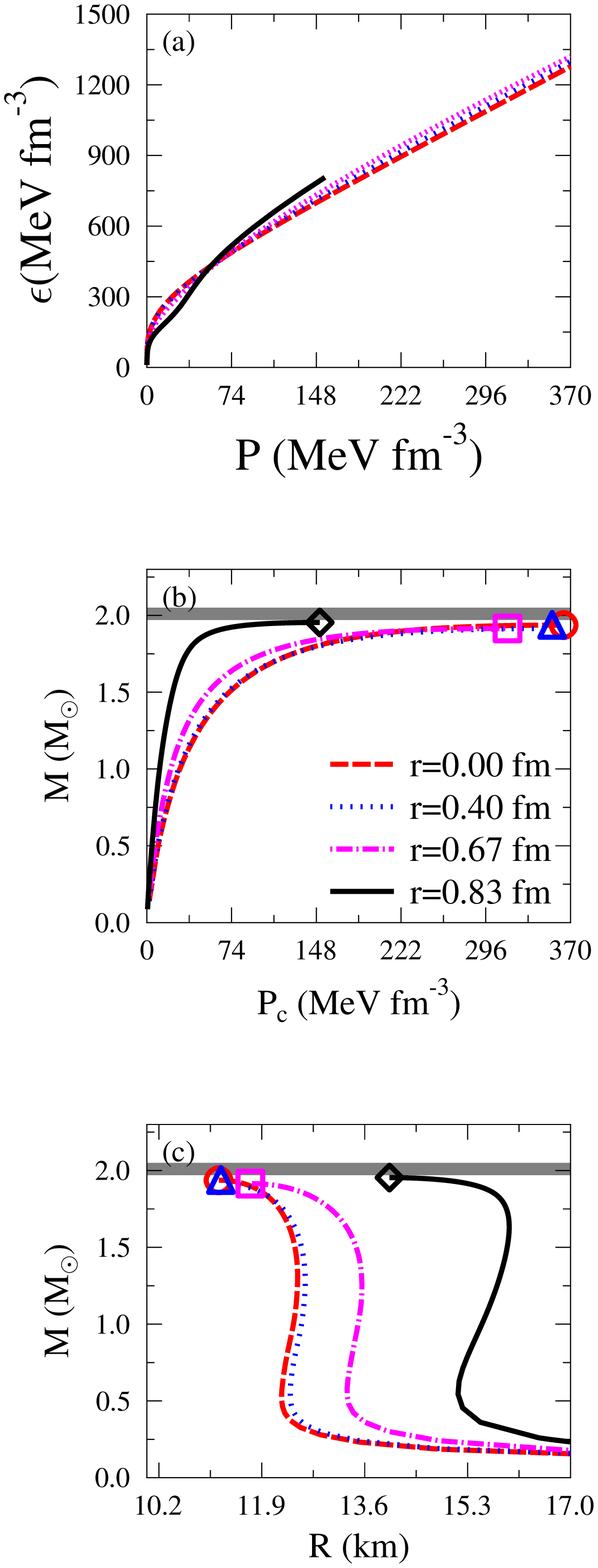, width=8cm}
\caption{(Color online) (a) Neutron star matter EOS, (b) mass as a function of the central pressure, and (c) mass-radius relation of the NS. Horizontal shaded bands in (b) and (c) are the pulsar mass constraint from Ref. \cite{Antoniadis13}.
  The circle, triangle, square, and diamond in (b) and (c) denote the points with the maximum mass.}
\label{fig:massrad}
\end{figure}

Figure~\ref{fig:massrad} displays the NS mass-radius, NS mass-central 
pressure, and NS matter EOS relations. From Fig.~\ref{fig:massrad}c
it appears that the nucleon radius $r$ does not significantly affect 
the maximum mass of NS. However, as the nucleon radius $r$ increases, the NS radius 
also increases. A more significant effect appears for $r \gtrsim 0.4$ fm. In the case of $r$ = 0.83 fm, the radius of 
canonical NS becomes very large, i.e., $R_{1.4 M_\odot} \approx 15.5$ km. 
The fact that the NS consisting of composite nucleons with large nucleon radius yields also a large NS radius can be understood by observing the 
relation between the NS central pressure $P_c$ and the NS mass $M$
as well as the  NS matter EOS shown in Figs.~\ref{fig:massrad}b and \ref{fig:massrad}a, respectively. 

From  Fig.~\ref{fig:massrad}b it is clear that for a large nucleon radius 
$r$ the NS mass $M$ increases faster as the central pressure $P_c$ increases. This phenomenon originates from the fact that the NS matter with relatively low pressure but large nucleon radius has a relatively stiffer EOS, whereas the EOS becomes relatively softer as soon as the pressure gets high.

\begin{figure}
\epsfig{figure= 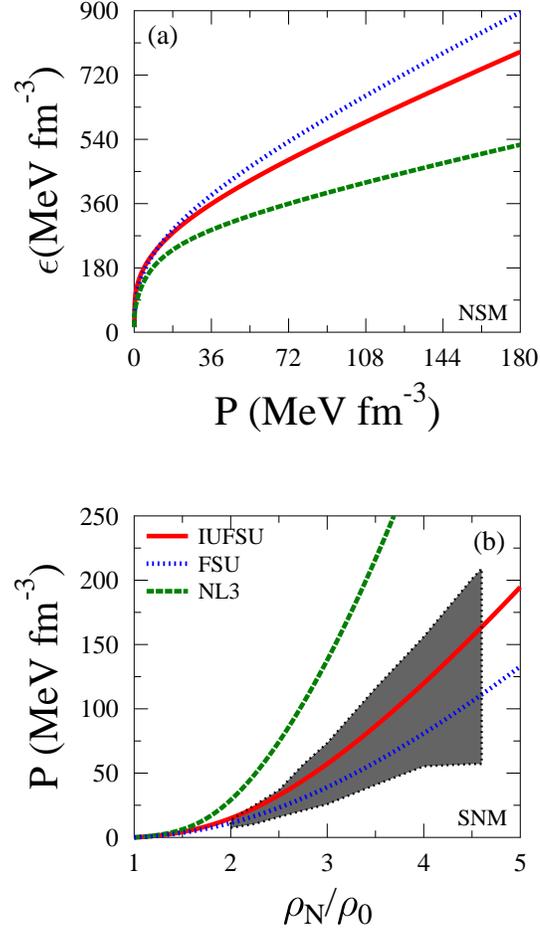, width=8cm}
\caption{(Color online) (a) Neutron star matter EOS and (b) SNM EOS of the standard IUFSU~\cite{Fattoyev2010}, NL3~\cite{LKR1997}, and FSU~\cite{TRP2005} parameter sets for comparison. }
\label{fig:nsm_STD}
\end{figure}

\begin{figure}
\epsfig{figure= 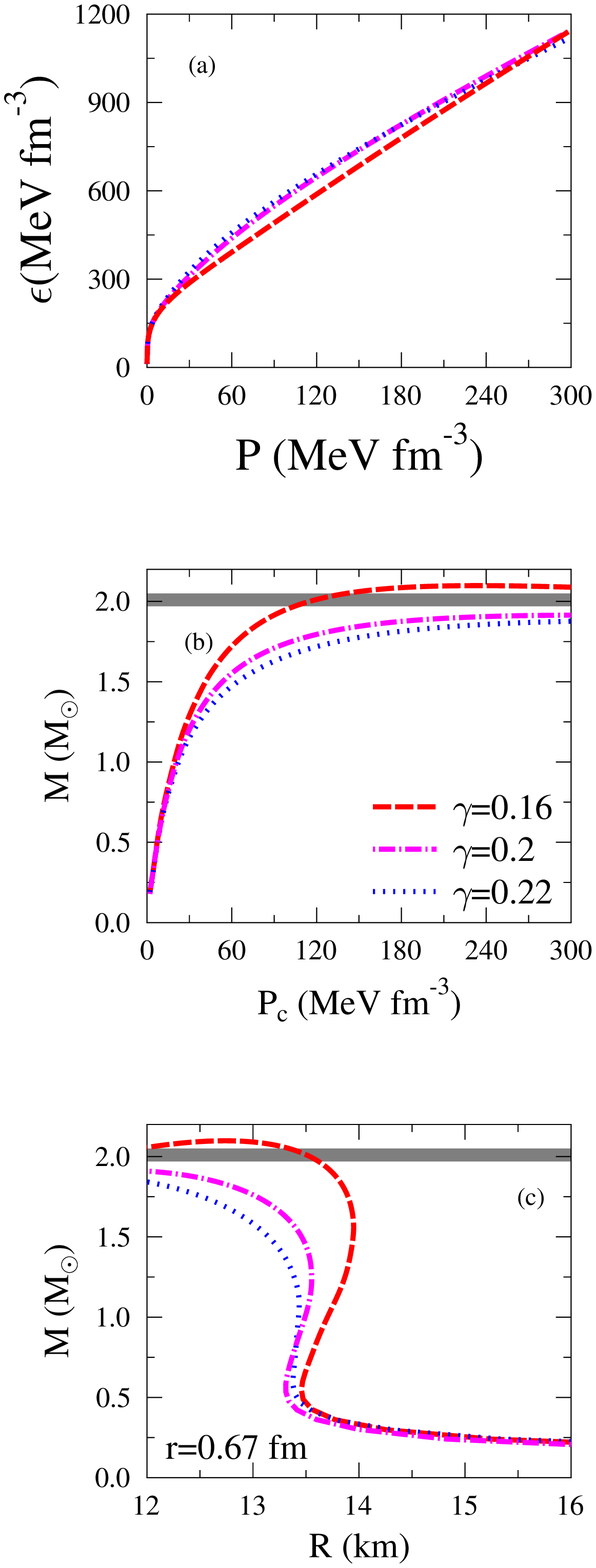, width=8cm}
\caption{(Color online) The influence of $\gamma$ on (a) NS matter EOS, (b) NS mass as a function of the central pressure, and (c) NS mass-radius relation for $r=0.67$ fm.}
\label{fig:massradr06}
\end{figure}

\begin{figure}
\epsfig{figure= 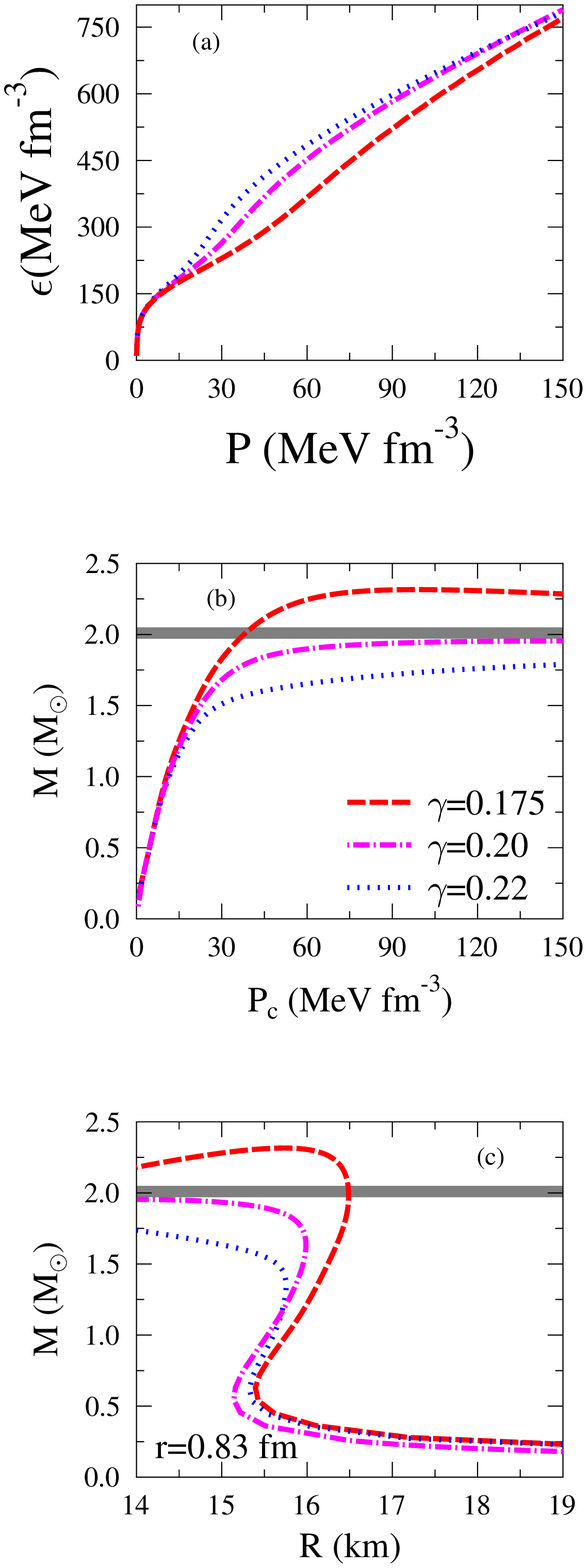, width=8cm}
\caption{(Color online) The influence of $\gamma$ on (a) NS matter EOS, (b) NS mass as a function of the central pressure, and (c) NS mass-radius relation for  $r=0.83$ fm.}
\label{fig:massradr083}
\end{figure}

Furthermore, at relatively high pressure the softness difference of the corresponding NS matter with difference $r$ is not too significant. This result is quite different from the one found in the case of point particle NS matter predicted by the standard parameter sets (IUFSU~\cite{Fattoyev2010}, NL3~\cite{LKR1997}, and FSU~\cite{TRP2005}), as shown in Fig.~\ref{fig:nsm_STD}, where the role of high pressure NS matter EOS of each corresponding parameter set is very significant. Therefore, the insensitivity of maximum mass and the strong dependence of the predicted NS radius on the free space nucleon radius originate mainly from the interplay between the contribution of relatively stiff EOS at low pressure and the contribution of relatively soft EOS at high pressure. To check the role of NS matter EOS at relatively low pressure, we plot the effect of $\gamma$ variation in the NS mass-radius relation for free space radii $r=0.67$ fm (Fig.~\ref{fig:massradr06}) and $r=0.83$ fm (Fig.~\ref{fig:massradr083}). It is obvious that the smaller $\gamma$ corresponds to the stiffer EOS at low density or low pressure. The significance of the effect depends on the value of nucleon  free space radius $r$ and it is obvious that the effect becomes more significant as the nucleon radius $r$ gets  larger. This result indicates that for this particular form of $r(\rho)$ [Eq.~(\ref{rrho})] the predicted NS maximum mass is also sensitive to the value of $\gamma$, whereas the predicted NS radius depends significantly on the free space nucleon radius $r$. This result corroborates the finding of Ref.~\cite{MS2013}. We also need to point out here that the forms of function $r(\rho)$ are not all causal. We have checked a number of different forms of $r(\rho)$ that are causal and mimic the behavior of  $r(\rho)$ as predicted by the quark meson coupling model~\cite{Panda2002,Aguirre2003}, and found that the larger $r$ corresponds to the larger NS radius. However, we still believe that the question of the existence of $r(\rho)$ that leads to a decreasing NS radii with increasing $r$ is still open. We defer this problem to the future investigation.

\begin{figure}
\epsfig{figure= 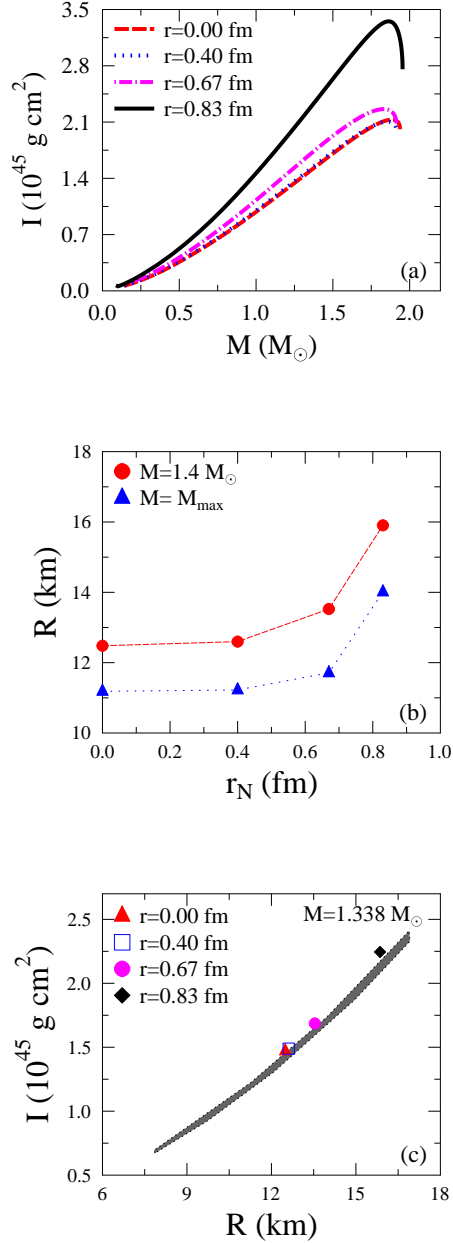, width=7cm}
\caption{(Color online) (a) Moment of inertia as a 
  function of NS mass, (b) the predicted radius of 
  canonical NS and the maximum NS mass as a function of the 
  nucleon radius, and (c) the NS moment of inertia as a function the NS radius for $M=1.338 M_{\odot}$. Shaded region in (c) 
  exhibits the approximated range of the moment of inertia obtained
  by Lattimer and Schutz~\cite{LatSch} for  $M=1.338 M_{\odot}$. 
  Note that in (b) the results obtained for 
  $M=1.338 M_{\odot}$ and $M=1.4 M_{\odot}$ are coincident. 
 }
\label{fig:rnvsR}
\end{figure}

In Fig.~\ref{fig:rnvsR}b, we show the non linear 
correlation between nucleon radius and the canonical NS radius 
$R_{1.4 M_\odot}$ as well as the maximum mass radius $M_{\rm max}$. 
These correlations emphasize that the effect of free space nucleon radius $r$ in 
matter is significantly imprinted in the NS radius $R$, especially for 
a relatively large $r$. Therefore, it is also interesting to 
see the effect of nucleon radius variation on the NS moment of inertia.

Figure~\ref{fig:rnvsR}a shows that the effect of larger 
NS radius $R$ on the moment of inertia $I$ 
becomes significant only for $r \gtrsim 0.7$ fm. 
It is clear that the moments of inertia $I$ calculated 
with $r = 0$ and $r$ = 0.83 fm 
differ significantly. In fact, the maximum $I$ in both cases differ
by $\approx 1.4\times 10^{45}$ g\,cm$^2$ or the corresponding ratio is $\approx 0.64$. Figure~\ref{fig:rnvsR}c shows the 
moment of inertia $I$ as a function of NS radius $R$ for 
$M=1.338 M_{\odot}$. The variation of $I$ as a result of the
variation of nucleon radius $r$ can be estimated by projecting the 
dependency of $R$ on $r$ as shown in Fig.~\ref{fig:rnvsR}b.
In the latter we can clearly see that the difference in $R$ 
turns out to be more than 1 km.  Therefore, if the uncertainty problem 
of the NS EOS had been solved and a precise measurement of the  
PSR J0737-3039A moment of inertia would exist in the future, 
the corresponding measurement could be used to provide a more stringent 
constraint on the free space radius of nucleon.

\begin{figure}
\epsfig{figure= 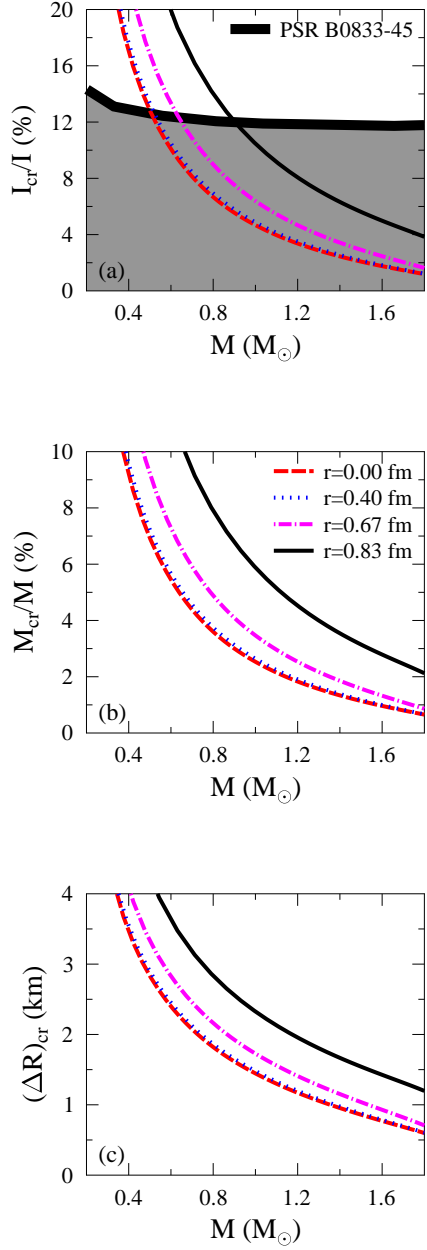, width=7cm}
\caption{(Color online)  (a) The ratio of the crust moment of inertia to the NS moment of inertia, (b) the ratio of the crust mass to the corresponding NS mass,  and (c) the thickness of the crust
  as a function of the NS mass. 
  Shaded area below the thick line in (a) shows 
  the excluded ratio from the pulsar timing data if the
  giant glitches in the Vela pulsar (PSR B0833-45)
  originate only from the NS crusts~\cite{Delsate2016}.
  }
\label{fig:crust}
\end{figure}

In addition, we also show the predictions for the NS crust properties,
such as the ratio of the crust mass $M_{\rm cr}$ to the NS mass $M$ as a 
function of $M$ in Fig.~\ref{fig:crust}a, 
the ratio of crust moment of 
inertia $I_{\rm cr}$ to the NS moment of inertia  $I$ as a function of 
$M$ in Fig.~\ref{fig:crust}b, as well as the crust thickness 
${(\Delta R)}_{\rm cr}$ as a function of $M$ in Fig.~\ref{fig:crust}c. 
It is obvious that the impact of free space nucleon radius $r$ 
in the NS crust properties is quite 
significant, i.e., increasing the radius $r$ will increase 
the thickness, mass and moment of 
inertia of the NS. The effect 
starts globally to appear at $r \gtrsim$ 0.4 fm. 
Compared to the one 
calculated with $r = 0$, the crust thickness ${\Delta R}_{\rm cr}$  
calculated with $r = 0.83$ fm differs by about 1 km, whereas the 
variance in the moment of inertia ratio $I_{\rm cr}/I$ of the two cases
is almost 50 $\%$. From Fig.~\ref{fig:crust}b we can also observe 
that the difference in the calculated $M_{\rm cr}/M$ of the two cases 
is nearly 40 $\%$. 

The findings reported in Refs.~\cite{Delsate2016} and ~\cite{PFH2014} are also
important to discuss here. By assuming that 
only neutron superfluid in the NS is responsible for the 
observed giant glitches, the authors of Ref.~\cite{Delsate2016} 
obtained a constraint on the $I_{\rm cr}/I$ of Vela pulsar (B0833-45),
which is shown in Fig.~\ref{fig:crust}a. 
By using a different mean field model than the one used here and 
considering the unified core-crust of EOS as well as  the crust superfluid 
contribution, they obtained from this constraint that the mass of 
the Vela pulsar would be $M\simeq 0.66 M_{\odot}$ at most. However 
the corresponding mass is too low and it is not compatible with 
the one predicted from the cooling simulation of Vela pulsar~\cite{PPP2015}. On the other hand, the authors of Ref.~\cite{PFH2014} have shown that by using the EOS analysis that predicts the neutron-skin thickness between 0.2 fm and 0.26 fm, the Vela glitches can be explained for NS with $M$ = 1.6 $M_\odot$ by taking the current uncertainties in NM $S$($\rho$) into account.
Finally, Fig.~\ref{fig:crust}a indicates that by using the IUFSU parameter 
set, but disregarding the unified core-crust EOS and 
the superfluidity in crust, the obtained EOS ($r=0$) predicts a 
Vela pulsar mass of  $\simeq 0.5 M_{\odot}$, but if the 
nucleon radius $r$ increases 
to 0.83 fm, then the predicted Vela pulsar mass increases to 
$\sim 0.9 M_{\odot}$. This estimate shows roughly that the 
effect of non zero free space nucleon radius could be also 
imprinted in the Vela pulsar glitches observation.

\section{SUMMARY AND CONCLUSION}
\label{sec_conclu}

We have studied the effect of free space nucleon radius on the NM 
and slowly rotating NS properties within an RMF model
with the inclusion of omega-rho mixing nonlinear term. Although we 
used the same density dependent radius as in our previous work, 
the present work utilized the  parameter sets of an RMF model with 
different values of composite 
nucleon radius $r$ obtained by adjusting the isoscalar and $\beta$ 
parameters. The adjustment was performed, so that the SNM 
properties at sub-saturation density 
coincide with the prediction of the IUFSU parameter set
\cite{Fattoyev2010}. We have obtained that the composite nucleons 
with free space radius $r \ne 0$ still produce the acceptable NM
properties in the vicinity of saturation density. 
Experimental data of SNM and PNM 
at moderate densities from Danielewicz {\it et al.} provide an 
upper limit to the free space nucleon radius, i.e., $r = 0.83$ fm. 
This value is close to the one obtained in the recent analysis of 
proton charge radius extraction from the proton scattering data
\cite{Higin2016,Madox2016}. In view of this, 
the imprint of nucleon radius in the NM could 
shed a new light on the proton radius puzzle.
We have also found that the effect of free space nucleon radius 
can be imprinted in the canonical NS radius as well as in the moment 
of inertia and crust properties of NS. 
By using the nucleon radius $r = 0.83$ fm we obtain a quite large canonical 
NS radius, i.e., $R_{1.4 M_\odot} \approx 15.5$ km, whereas the 
corresponding impact on the moment of inertia $I$ becomes 
sizable for $r \gtrsim 0.7$ fm. In the crust properties, 
the effect starts to appear even at $r \gtrsim$ 0.4 fm. The latter
could have an impact on the observation of Vela pulsar glitches.


\section*{ACKNOWLEDGMENT}
The work of A.S. and T.M. was partly supported by the University of Indonesia.



\end{document}